\documentclass[12pt]{iopart}

\usepackage{iopams}
\usepackage{graphicx}% Include figure files
\usepackage{dcolumn}% Align table columns on decimal point
\usepackage{bm}% bold math
\usepackage{rotating}
\usepackage{multirow}

\begin{document}

\title{Empirical confirmation of creative destruction from world trade data}

\author{Peter Klimek$^1$, Ricardo Hausmann$^{2,3}$, Stefan Thurner$^{1,3,4,*}$}

\address{$^1$Section for Science of Complex Systems; Medical University of Vienna; 
Spitalgasse 23; A-1090; Austria\\ 
$^2$Center for International Development and Harvard Kennedy School, Harvard University, Cambridge, MA 02138; USA\\
$^3$Santa Fe Institute; 1399 Hyde Park Road; Santa Fe; NM 87501; USA\\
$^4$IIASA, Schlossplatz 1, A 2361 Laxenburg; Austria}

\ead{$^*$stefan.thurner@meduniwien.ac.at}

\begin{abstract} 
We show that world trade network datasets contain empirical evidence that the dynamics of innovation in the world economy follows indeed the 
concept  of {\it creative destruction}, as proposed by J.A. Schumpeter more than half a century ago. 
%At the same time the ??? can be ruled out. 
National economies can be viewed as complex, evolving systems, driven by a stream of appearance and disappearance of goods and services.
Products appear in bursts of creative cascades.
We find that products systematically tend to co-appear, and that product appearances lead to massive disappearance events of existing products in the following years.
The opposite -- disappearances followed by periods of appearances --  is not observed.  
%To our knowledge this is the first 
This is an empirical validation of the dominance of cascading competitive replacement events on the scale of national economies, i.e. creative destruction.
We find a tendency that more complex products drive out less complex ones, i.e. progress has a direction.
Finally we show that the growth trajectory of a country's product output diversity can be understood by a recently proposed
evolutionary model of Schumpeterian economic dynamics.
 \end{abstract}
 
\maketitle

Joseph A. Schumpeter held that the key mechanism of economic development is radical innovation \cite{schumpeter1942, Thurner10}. 
In his view the Walrasian economic equilibrium is continuously disturbed by actions of entrepreneurs, introducing novel goods and services in the market.
These innovations may replace existing goods and services and thereby impact related industries.
If this happens as a cascading process it is called {\it creative destruction}.
Current colloquial examples of how once market-dominating companies lose their position due to creative destruction include instant photography or printed newspapers
in light of the age of digitalization.
But is this only the driving force behind major shifts in industrial production and long-term business cycles, or does it also condition economic change on much shorter time-scales?

This question has been addressed using firm entry and exit dates \cite{Bartelsman04}, job creation and destruction rates \cite{Caballero96}, 
within specific countries \cite{Aghion08} or within specific industries \cite{Tripsas97}. 
These works capture the complex interplay between factor re-allocation and productivity growth in existing economic sectors.
However, little is added to our understanding of the impact of emerging industries on the development of already established parts of the national economy.
If a new industrial branch emerges, how does this impact other economic sectors?

Traces of the creation of unprecedented industries can be observed using world trade data.
We study the dynamics of the diversity of export products and show that the process through which it changes follows the patterns of creative destruction.
The set of products a given country exports reveals the presence of the nontradable inputs or capabilities that the products require for their production
 (e.g. specific productive knowledge, infrastructure, legal system, labor skills, regulations, etc.) \cite{Hidalgo09, Hausmann11a}.
In this view capabilities are elementary building blocks and each product requires a combination of them to be manufactured.
A change in the diversity of a country's product basket indicates a change in its set of capabilities.
New capabilities may lead to new products and the abandonment or substitution of already existing products or capabilities.
Products requiring one of these abandoned capabilities as input will then disappear,
 while simple products that require only subsets of capabilities of more complex products may disappear as they are unable to compete for these inputs  
 -- creative destruction at work.

In this article we show that typically within a country clusters of products appear simultaneously in bursts.
We interpret such a burst as the acquisition of a novel capability which is required as input for each of the newly appearing products.
In the years following creative bursts there is an increased chance that already existing products will cease to be exported,
that is, emerging industries effectively push out the old ones.
The novel products tend to be more sophisticated and complex than the disappearing products, i.e. progress has a direction.
Interestingly, we do not observe a higher probability of product appearances following disappearance events of products.
This means that the canonical mechanism of filling "market niches" is not a significant feature of the data.
We confirm empirically that creative destruction, cascades of competitive replacement, plays an important role in the development of national economies.

We apply a recently proposed Schumpeterian diversity dynamics model \cite{Thurner10} to account for the evolution of the product diversity of countries.
In this model entrepreneurs use and combine available capabilities to create novel goods and services and substitute them for already existing ones.
The result is a model economy in a self-organized critical state, characterized by creative and destructive co-evolutionary avalanches.
The changes in product diversity and the distribution of product appearances and disappearances observed in world trade data are well explained by this model.
We base our results on the World Trade Flows database compiled by the National Bureau for Economic Research \cite{NBER}.

\section{Results}

\subsection{Diversity dynamics}
The World Trade Flows database \cite{NBER} contains exports of approximately 200 countries over the years 1984-2000 in about 800 product categories (4-digit SITC rev.2 classification).
We include only those countries in the analysis which have a population of at least 1.2 million people and total exports of at least 1 billion USD, for more information see Text S1 in the SI.
We label the countries contained in this database by $c$, the product category by $p$ and the year by $t$.
The export values $x(p,c,t)$, denominated in USD are then extracted.
As a diversification measure of country $c$ in year $t$ the number of products with nonzero export values is used, that is, $D_c(t) = \sum_p \mathrm{sgn}(x(p,c,t))$.
The total number of countries is denoted by $N_C$, the number of products by $N_P$.

\begin{figure}
\begin{center}
 \includegraphics[width=120mm]{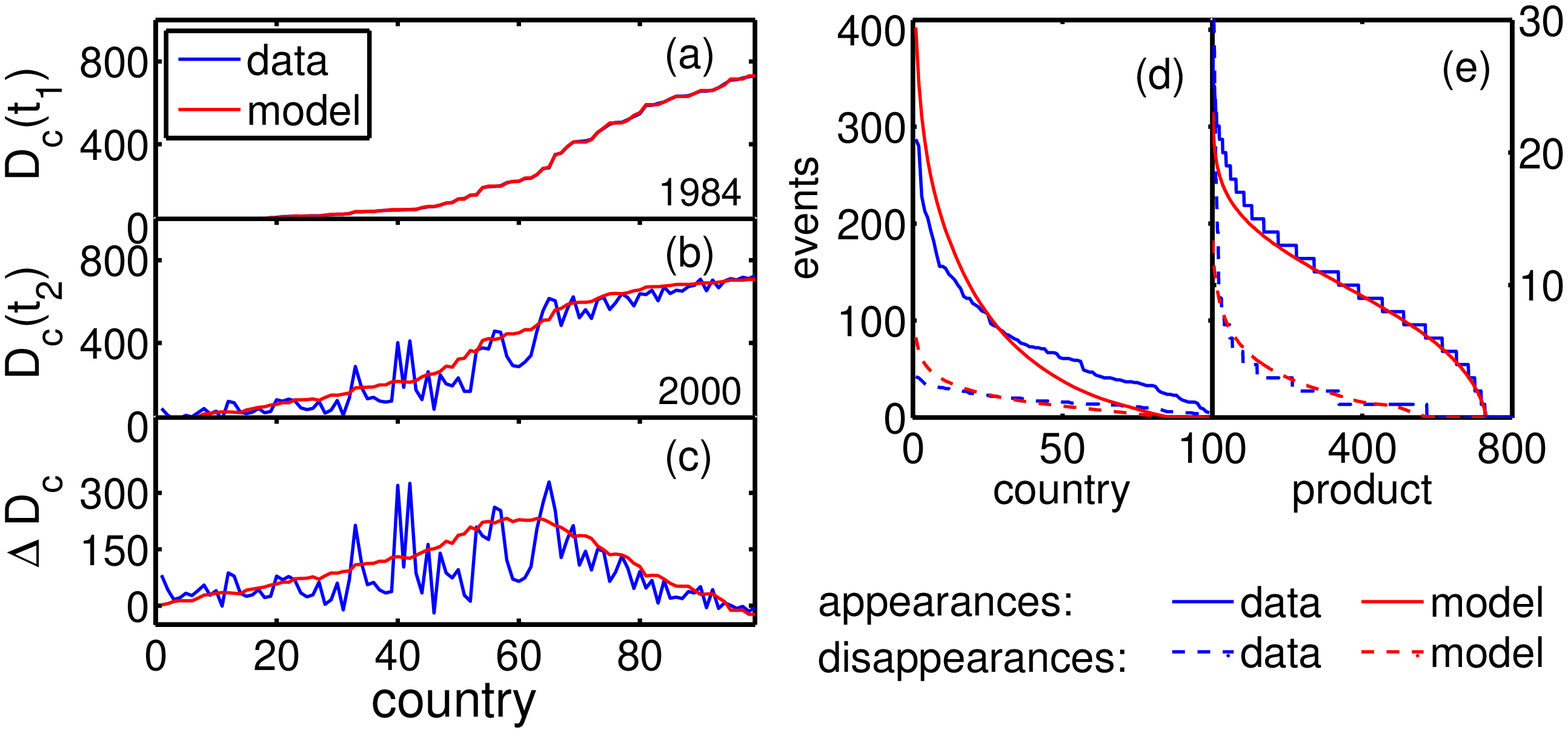}
\end{center}
 \caption{Countries' export product diversities for (a) 1984, (b) 2000 and the net change in diversity between these years (c) are shown for world trade data (blue lines) and the Schumpeterian diversity dynamics model (red). There is a general tendency towards increased diversity which is strongest for countries having initially an intermediate product diversity. The results for diversity change are obtained by using only the subset of countries having non-zero export values in each year. The number of (dis)appearance events for each (d) country and (e) product are shown in the panel to the right for data and model. The model does not only reproduce the general trend of increasing diversity, but also the detailed profile of how many appearance and disappearance events could be observed for different countries and products. Note that in (a)-(c) countries are ranked by their initial diversity, whereas in (d) they are ranked by their number of events.}
 \label{Schump}
\end{figure}

The change of diversity of product exports in world trade data over a timespan of 16 years is shown in Fig.\ref{Schump}(a)-(c).
For each country index $c$ the diversity $D_c(t)$ is shown for two different years, $D_c(t_1=1984)$ in Fig.\ref{Schump}(a) and $D_c(t_2=2000)$ in Fig.\ref{Schump}(b).
The net change in diversity for each country between 1984 and 2000, $\Delta D_c = D_c(t_2=2000)-D_c(t_1=1984)$ is shown in Fig.\ref{Schump}(c).
Values from trade data are shown in blue and are compared to results of the Schumpeterian diversity dynamics model (in red, to be explained below).
There is a general trend towards increased diversity.
Countries with a relatively low or high diversity in 1984 show smaller fluctuations in diversity than countries with intermediate initial diversity.
That is, countries with a low diversity tend to stay poorly diversified, fully diversified countries stay fully diversified.
In between there is a regime of transition countries, some of them showing explosive growth in terms of their economic diversity.
We show later that these observations can be described through the onset of a 'creative phase transition' \cite{Thurner10}, a distinguishing property of Schumpeterian diversity dynamics.
Let us now empirically investigate the process by which the countries' product export diversities change.

Let $A(p,c,t)$ be a product indicator function for the appearance of product $p$ in country $c$ between year $t-1$ and $t$,

\begin{equation}
A(p,c,t) = \left\{
  \begin{array}{l l}
    1 & \quad \mathrm{if } \ x(p,c,t-1)\leq\theta \ \mathrm{ and} \ x(p,c,t)>\theta\quad,\\
    0 & \quad \mathrm{otherwise}\quad.\\
  \end{array} 
 \right.
\label{AppEv}
\end{equation}
Similarly the indicator function for a product disappearance is
\begin{equation}
D(p,c,t) = \left\{
  \begin{array}{l l}
    1 & \quad \mathrm{if } \ x(p,c,t-1)> \theta \ \mathrm{ and} \ x(p,c,t)\leq \theta\quad,\\
    0 & \quad \mathrm{otherwise}\quad,\\
  \end{array} 
 \right.
\label{DisAppEv}
\end{equation}
with a threshold value set to $\theta=$ 100,000 USD.
For more details see Fig. S1
Fig.\ref{Schump}(d) and (e) show the distribution of (dis)appearance events $A$ ($D$) per country $c$ and product $p$ for world trade data and the Schumpeterian diversity dynamics model.
They are far from being homogeneously distributed.
For example, the number of appearance events per country varies between more than 300 and almost zero across different countries. 
Further there is a substantial number of products appearing in, say, five countries or less, while others appear in almost one third of all the countries studied.
The model captures the functional form of these distributions.
The number of appearances clearly exceeds the number of disappearances, consistent with the general trend towards higher diversity.

\subsection{Co-occurrence analysis}
Creative destruction can be described as a process started by introducing novel goods or services in a national economy.
If successful, this stimulates the market introduction of related  goods, complimentary to the newly introduced ones.
Thereby a novel cluster of inter-related products can form.
This cluster may render existing parts of the national economy obsolete.
If this process is actually at work within national economies, one would expect two empirical facts to hold.
To pin down this very process empirically, one needs to measure two features, (i) products enter the market in creative bursts, i.e. not incrementally and (ii) the appearance of goods tends to foster the disappearance of other products.
We quantify this with two different product indicators referred to as 'Schumpeterian Product Indicators' ($SPI$). 
The first is introduced to assess co-occurrences of product (dis)appearances and is denoted $SPI^0_{XY}(p)$.
Here $X$ and $Y$ stand for any combination of events $A,D$ for product $p$. The second $SPI$ measures if an appearance or disappearance event of products at $t$ is correlated to an event of another product within the following $\tau$ years and is denoted by $SPI_{XY}^{(\tau)}(p)$.

The marginal appearance (disappearance) frequency $P_A(p)$ ($P_D(p)$) for each product is given by 
\begin{equation}
  \begin{array}{l l l}
    P_A(p) & = & \sum_{c,t} A(p,c,t)\quad,\\
    P_D(p) & = & \sum_{c,t} D(p,c,t)\quad.\\
  \end{array} 
\label{MargProbsP}
\end{equation}
The number of co-appearances $P_{AA}(p,q)$ of products $p$ and $q$ across all countries and times is,
\begin{equation}
P_{AA} (p,q) = \sum_{c,t} A(p,c,t) A(q,c,t) \quad.
\label{PAA}
\end{equation}
The co-occurrence\footnote{By co-occurrence we will refer to any of the four possible pairs of events for two products $p$ and $q$, that is (dis)appearance in $p$ together with (dis)appearance in $q$.} statistics $P_{AD}(p,q)$ and $P_{DD}(p,q)$ are obtained by appropriately substituting disappearance events $D(p,c,t)$ for $A(p,c,t)$ in Eq. \ref{PAA}.
Obviously products with a relatively high number of appearances will also be more likely to co-appear.
Consequently whenever one compares co-occurrence statistics of two different pairs of products one has to correct for this bias. 
A simple way to do this is to compare the number of measured co-occurrences of two products to their marginal (dis)appearance frequencies given in Eq. \ref{MargProbsP}.
For this we define the pairwise conditional co-occurrence measure as $\tilde P_{AA}(p,q) \equiv \frac{P_{AA}(p,q)}{\max \left[P_A(p), P_A(q) \right]}$.
The idea is that given two products $p$ and $q$ we take the product with the higher appearance probability, say $p$, and measure how often $q$ appears conditional on an appearance event of $p$ in the same country.
To quantify in how many co-occurrence events a given product participates, one can define an $SPI$ for simultaneous appearances $SPI_{AA}^0 (p)$ by 
\begin{equation}
SPI_{AA}^0 (p) = \frac{1}{\mathcal N}  \sum_q \tilde P_{AA}(p,q) \quad.
\label{SPI0}
\end{equation}
where $\mathcal{N}$ is a normalization factor given by $\mathcal{N} = (N_P -1) N_C$ guaranteeing that the index lies within the range $(0,1)$ and is thus comparable across datasets of different sizes.
It is straight forward to define $SPI$s for other pairs of events $XY \in \{AD,DA,DD\}$.

We proceed to quantify to what extent an appearance event in year $t$ is related to disappearance events at a later time $t'>t$.
Assume a maximal lag of $\tau$ years between the two events, $t < t' \leq t+\tau$.
If not denoted otherwise, we work with a value of $\tau=3$ years.
A simple measure is the number of all appearances of $p$ which are followed by disappearances of $q$ within the next $\tau$ years in the same country and summed over all countries, i.e. $P^{(\tau)}_{AD}(p,q) = \sum_{c,t} \sum_{t'=t+1}^{t+\tau} A(p,c,t) D(q,c,t')$.
This number can be compared to the count of appearances of product $q$ followed by disappearances of $p$, i.e. we exchange the roles of $p$ and $q$.
If there an asymmetry emerges when exchanging $p$ and $q$ in $P^{(\tau)}_{AD}(p,q)$,
this indicates that $p$ appearing before $q$ disappears is observed more often than the other way around.
In this spirit, for each product $p$ a time-lagged $SPI$ is defined as 
\begin{equation}
SPI_{AD}^{(\tau)}(p) = \frac{1}{\mathcal N} \sum_q \left[ P^{(\tau)}_{AD}(p,q) - P^{(\tau)}_{AD}(q,p) \right]
\quad,
\label{SPIt}
\end{equation}
This index is within the range $(-1,+1)$.
Intuitively, if one thinks of the appearance of $p$ followed by disappearance of $q$ within a country as a replacement, a positive value of $SPI_{AD}^{(\tau)}(p)$ means that, on average, $p$ replaces more often any other product $q$ than $p$ itself is replaced by $q$. 
The higher the $SPI_{AD}^{(\tau)}(p)$ value for $p$, the higher the tendency that $p$ can act as a substitute for other products.

\subsection{Products appear in bursts}
\begin{figure}[tbp]
 \begin{center}
 \includegraphics[width=140mm]{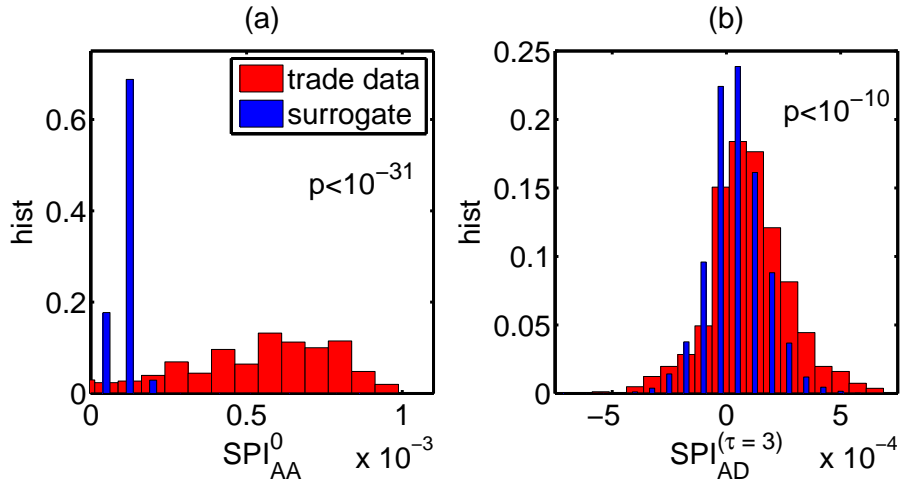}
  \end{center}
 \caption{Histograms for Schumpeterian Product Index (a) $SPI^0_{AA}(p)$ for co-appearances $AA$, and (b) $SPI^{(\tau)}_{AD}(p)$ for $\tau=3$ years. $SPI$s are shown for the surrogate data (blue) and the trade data (red). The distributions for the trade data are significantly shifted to the right when compared to the surrogate data. The distribution in (a) suggests that products appear in bursts, while (b) implies that appearing products tend to drive other products from the market.}
 \label{SPIs}
 \end{figure}

\begin{figure}[p]
 \begin{center}
 \includegraphics[width=122mm]{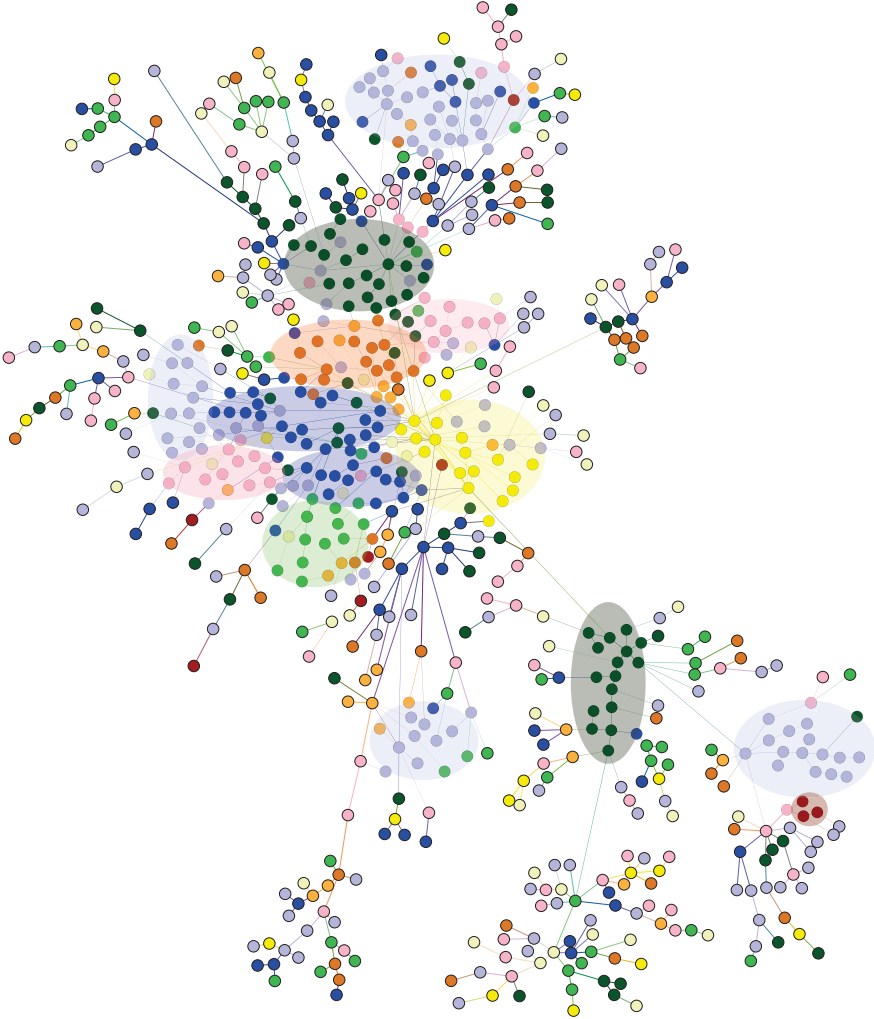}\\
 \includegraphics[width=120mm]{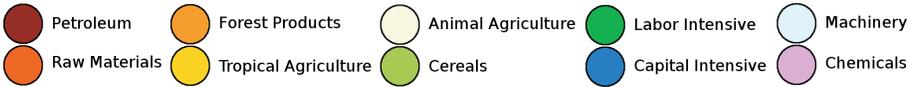}
 \end{center}
 \caption{Maximum spanning tree for the network given by the co-occurrence measure $\tilde P_{AA}(p,q)$. Some of the clusters of co-appearing products are highlighted as guides to the eye. It is suggested that these clusters of products require a common capability before they can be exported. Once a country acquires all of these capabilities, a creative burst of novel products which require them as input may be the consequence.}
 \label{MST}
\end{figure}

In Fig. \ref{MST} we show the maximum spanning tree for products obtained from the pairwise conditional co-appearance measure, $\tilde P_{AA}(p,q)$.
The clusters highlighted there point at a non-random structure in product co-appearance dynamics.
To make this structure explicit we compare $SPI^0_{AA}(p)$ values for trade and the surrogate data, see histogram in Fig. \ref{SPIs}(a). 
A huge difference between trade and surrogate data is found.
This indicates the presence of strong temporal correlations between individual product appearances.
Products appear simultaneously in bursts. A way to interpret this is that the co-appearing products
require a common capability to be manufactured. 
When a national economy acquires this capability the corresponding cluster of products can appear.
In this sense the observation of a newly emerging cluster of inter-related products serves as a proxy for a country's development of a novel capability.
This process is not unique within a given country, it can be observed in a substantial fraction of them (since the indicator values are averages calculated over more than one hundred countries).
This may hint at common patterns in the diversification trajectory of individual national economies.

\subsection{Creative destruction at work}
There is a tendency for specific products to substitute or replace others. 
This can be seen as deviations in the distribution of indicator values $SPI^{(\tau)}_{AD}(p)$ between world trade and surrogate data, see Fig. \ref{SPIs}(b).
A high value of $P^{(\tau)}_{AD}(p,q)-P^{(\tau)}_{AD}(q,p)$ corresponds to the dominance of the pattern: 'product $p$ appears and $q$ disappears later in the same country' in a large number of them independently.
This is a direct fingerprint of creative destruction at work, i.e. emerging industries pushing out the old. 
It is interesting to note that the opposite process -- the disappearance of a product is followed by the appearance of another one -- is not observed to a significant extent, see Figs. S2 and S9.
For the $DA$ case surrogate and data $SPIs$ are practically identical.
To measure the statistical significance of deviations from the surrogate data we formulate the null hypothesis that both surrogate and trade data are drawn from a normal distribution with the same mean.
The $p$-value using the alternative hypothesis that the trade data shows a higher mean than the surrogate data is computed 
for all four possible combinations: $AA$, $AD$, $DA$, and $DD$.
The results are listed in Tab. 1, together with $p$-values from various robustness tests as described in the SI.
This series of tests confirms that the results above are not a consequence of trivial effects such as fluctuations in the trade records (case '$\theta$ = 200k USD'), the artificial creation of product categories due to changes in reporting schemes ('all products with positive exports') or the transition out of communism of former Soviet countries ('Excl. FSU'), see SI.
Results are also robust with respect to changes in the time-period ('1989-2000'), the choice of the maximal lag $\tau$ ('$\tau=5$','$\tau=7$') or the underlying dataset ('UN Comtrade data').

\subsection{Progress has a direction}
Is it true that more complex products replace less complex ones? Do products produced by richer countries replace products associated with poorer ones?
To elucidate this, we quantify the change in complexity and income levels associated with individual $AD$ processes.
The creative destruction dynamics is dominated by products with a high value of $KI(p)$ or $XI(q)$.
$\Delta PCI(p,q,\tau)$ and $\Delta PRODY(p,q,\tau)$ are thus measured for each $AD$ process with $\tau=3$,
 where the appearing (disappearing) product has one of the hundred highest values of $KI(p)$ ($XI(q)$).
A positive $\Delta PCI$ indicates an increase in economic complexity of the country where this $AD$ process was observed. 
A positive $\Delta PRODY$ indicates that the country is upgrading its exports towards those made by richer countries.
Histograms for $\Delta PCI$ and $\Delta PRODY$ are shown in Fig. \ref{DPCI}. There is a clear skew to the positive side for both quantities.
Products with a higher tendency to be pushed from the market have in general lower economic complexity than products from appearing industrial branches.

One could imagine that this effect is stronger in emerging economies than in mature ones.
To test this, one can compare $\Delta PCI$ and $\Delta PRODY$ distributions for different economies by aggregating the countries into seven regions (Latin America and the Carribean, East-Asia and the Pacific, Middle-east and North-Africa, Sub-Saharan Africa, South-Asia, Western Europe and Northern America, Eastern Europe), as done in the SI.
A clear development towards higher economic complexity can be seen for Latin America, Eastern Europe and East Asia \& Pacific. The other regions do not display such a strong trend. 
Western Europe and Northern America are already almost fully diversified and stay that way, whereas e.g. Sub-Saharan Africa stays in the low diversity regime.
These results are discussed in more detail in the SI.

\subsection{Which products are the killers?}
\begin{figure}[tbp]
  \begin{center}
  \includegraphics[width=140mm]{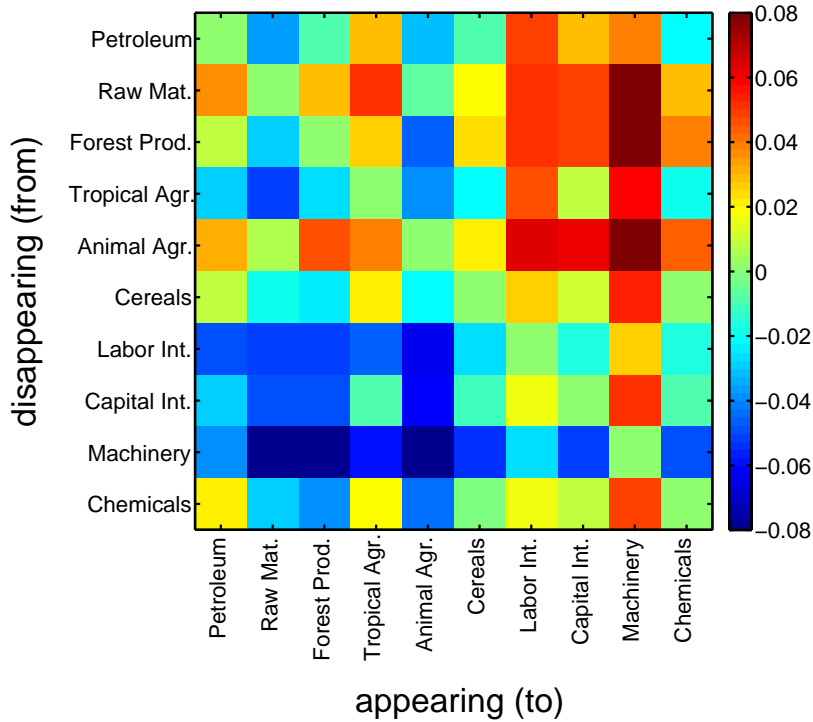}
   \end{center}
  \caption{Transition matrix $\Pi$ between Leamer classes. For each pair of classes $(l_i,l_j)$ it is measured how often an appearance event of a product from class $l_i$ is observed followed by a disappearance of a product from class $l_j$, compared to the opposite direction. Rows are indexed by the appearing Leamer classes $l_i$, columns by disappearing ones $l_j$. A positive value indicates an excess in creative destruction processes between the two classes (from row index to column index). The matrix is by construction anti-symmetric. There is a clear tendency of appearing labor and capital intensive products, as well as machinery and chemicals. Cereals or agricultural products tend to disappear.}
   \label{LeamerFlow}
  \end{figure}

Let us get a clearer picture about the products driving the creative destruction process.
Are there Leamer classes whose products appear more often than others? Which types of products disappear instead?
We consider the matrix $P^{(3)}_{AD}(p,q)$. Each product can be assigned to one of ten Leamer classes, labeled by $l_i$, $i \in \{1, \dots 10\}$.
One can obtain a measure for $AD$ processes where products from class $l_i$ appear and products from $l_j$ disappear by computing the mean value of $P^{(3)}_{AD}(p,q)$ over all products from the respective Leamer classes.
To this end $L(l_i,l_j) =  \frac{1}{\mathcal N} \langle P^{(3)}_{AD}(p,q) \rangle_{p \in l_i, q \in l_j}$ is defined.
As before, we are only interested in the 'net flow' of $AD$ processes, the antisymmetric Leamer transition matrix, $\Pi=L-L^{T}$.
This is basically the same measurement strategy as discussed for $P^{(\tau)}_{AD}(p,q)-P^{(\tau)}_{AD}(q,p)$, however aggregated to the level of Leamer classes.
 The matrix is shown in Fig. \ref{LeamerFlow}.

\begin{figure}[t]
 \begin{center}
 \includegraphics[width=140mm]{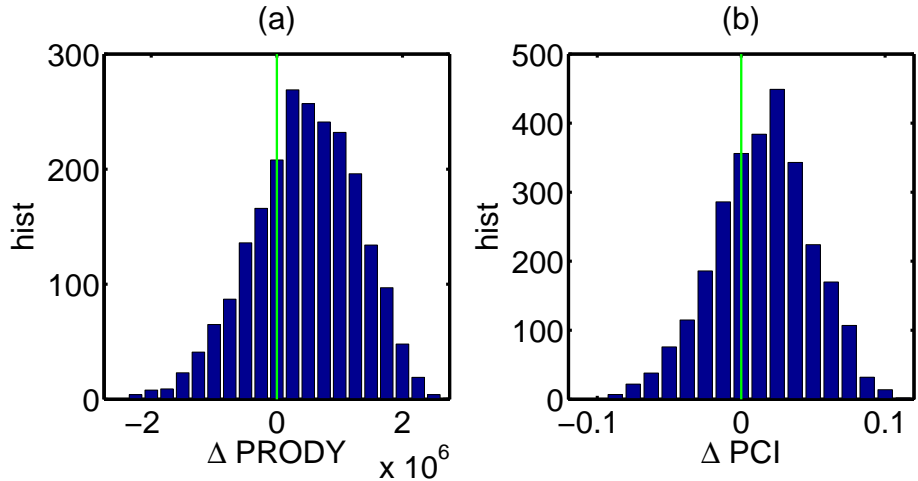}
  \end{center}
 \caption{Histograms for change in (a) income level $PRODY$ and (b) product complexity $PCI$ associated with individual creative destruction processes. There is a skew to the positive side in both of them, indicating that national economies are typically restructured  into the direction of more complex and higher income products.}
 \label{DPCI}
\end{figure}

A positive value of $\Pi(l_i,l_j)$ indicates that an appearance event of a product from class $l_i$ followed by disappearance of a product from $l_j$, is more often observed than an appearance from $l_j$ followed by disappearance in $l_i$. 
Given that $AD$ processes convey information about how national economies are re-structured over time, positive values in $\Pi$ indicate which parts of the economy are abandoned or disappear because of which other parts.
There is a clear trend. Capital and labor intensive products, as well as machinery and chemicals appear much more often than they disappear. Agricultural products and cereals tend to disappear. 
This intuitively confirms that markets re-structure themselves towards higher economic complexity as observed in Fig. \ref{DPCI}.
Note that these observations can not be described by diffusion or migration processes of e.g. production facilities or capabilities from one country to another.
Each measurement of appearances followed by disappearances takes place within a single country. 
A product appears within a country and another product disappears later within the same country.

\begin{table}[b]
\caption{$p$-values for $SPI^{(\tau)}_{XY}$ histograms from trade vs. surrogate data. The column '$\tau = 3$' lists the results as described in the main text. In addition, robustness tests where conducted with results listed in separate columns. '$\theta=$ = 200k USD' uses the same set of countries and a threshold of 200.000 USD below which trade flows are ignored. In 'all products with positive exports' all products are included which have positive world exports in each year of the analysis. The column '1989-2000' decreases the number of years included in analysis. Results excluding the FSU and CEE are listed in 'Excl. FSU'. The maximal lag $\tau$ is then increased to $\tau=5, 7$. The last column reports results using the UN ComTrade dataset, as described in the SI.}
 \label{TabPVal} 
\begin{tabular}{lcccccccc} 
\hline
 & \parbox{1.1cm}{$\tau=3$ }& \parbox{1.1cm}{\small $\theta$ = 200k USD}  & \parbox{1.1cm}{\scriptsize all products with positive exports} & \parbox{1.1cm}{\scriptsize cleaned data 1989-2000} & \parbox{1.1cm}{ \scriptsize Excl. FSU} & \parbox{1.1cm}{$\tau=5$} & \parbox{1.1cm}{$\tau=7$} & \parbox{1.1cm}{\scriptsize UN ComTrade data}  \\ \hline 
AA  & 0.98 & 1.00 & 0.93 & 0.96 & 1.00 & 0.94 & 0.84 & 1.00  \\
DD & 0.95 & 0.89 & 0.77 & 1.00 & 0.94 & 0.93 & 0.99 & 1.00 \\
AD &{\boldmath $<10^{-10}$}&{\boldmath $<10^{-6}$}&{\boldmath $0.0072$}&{\boldmath $<10^{-7}$}&{\boldmath $<10^{-5}$}& {\boldmath $<10^{-7}$}& {\boldmath $<10^{-3}$}& {\boldmath $<10^{-3}$}  \\
DA  & 0.02 & 0.11 & 0.53 & 0.01 & 0.40 & 0.08 & 0.12 & 0.081  \\
  \hline
\end{tabular}
\end{table}

\subsection{A simple model}

\begin{figure}[tbp]
  \begin{center}
  \includegraphics[width=80mm]{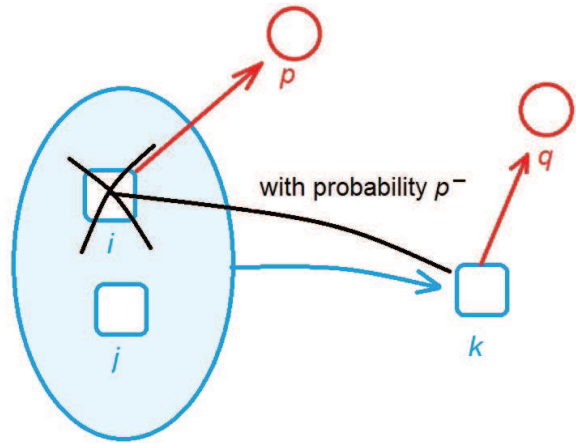}
   \end{center}
  \caption{Illustration of the Schumpeterian diversity dynamics model. Capabilities are represented by blue squares, products by red circles. 
A country can acquire capability $k$ by combining  other capabilities $i$ and $j$.
Each product requires a set of capabilities to be produced, e.g. product $p$ requires $i$ and $q$ requires $k$.
There is a chance that a novel product $q$ may act as a substitute for a product $p$ which is made from a subset of capabilities required for $k$.
In this case $k$ may effectively 'destroy' its preceding capability $i$.}
 \label{Scheme}
  \end{figure}

A recently introduced Schumpeterian diversity dynamics model \cite{Thurner10} centers around the assumption that countries have an evolving set of capabilities which firms combine to manufacture products, see SI.
The economic state of a country is represented by two (high-dimensional) binary vectors, one indicating whether a country has a given capability, the other indicating whether a country exports a given product.
These vectors evolve over time in each country.
Capabilities can be acquired through entrepreneurs who combine existing capabilities. 
For example, an existing production infrastructure and a certain type of knowledge stock can be combined to acquire an upgraded production facility.
This scheme is shown in Fig. \ref{Scheme}, where capabilities are represented by blue squares.
In this case capability $k$ is a combination (blue ellipse) of $i$ and $j$.
The model capability assumes that if a country has capability $i$ and $j$, it will acquire $k$ in the next time-step.

Each product requires a set of capabilities in order to be produced. 
In the model a country exports a product if it has {\it all} necessary capabilities.
The simplest possible case (one capability needed for one product) is shown in Fig. \ref{Scheme} where products are represented by red circles. 
Product $p$ requires capability $i$, product $q$ requires $k$.
As soon as a country acquires $k$, it will start to report exports in $q$.
Since one capability may be required for more than one product, all of these products may co-appear with $q$.

Capabilities can also be lost or abandoned.
There is a chance that $q$ may act as substitute for $p$. In this case the acquisition of capability $k$ renders $i$ obsolete.
With some probability $p^-$ capability $i$ is thus abandoned or destroyed because of $k$.
As a consequence $p$ will then no longer be exported.

In the model it is further assumed that the rules of how capabilities can be combined, substituted and used as inputs for products are identical in each country.
Economies only differ by their initial diversity of products (from which the implied diversity of capabilities is calculated).
The model is iterated until the number of model (dis)appearances matches the number of events observed in the trade data.
Fig. \ref{Schump} shows a comparison of (a) initial and (b) final product diversity, (c) net diversity change and the number of appearances for each (d) country and (e) product for model and trade data.
The complex observed patterns of creative destruction can be explained by the almost embarrassingly simple process of recombining and substituting capabilities.
The crucial feature of the model is the existence of a creative phase transition with a position depending on the initial diversity \cite{htk1, htk2}.
If it is below a certain threshold, there are not enough capabilities available to find successful combinations of them.
If a country is at, or above this threshold the creative destruction process kicks in, restructuring the market \cite{Thurner10}. 
For a complete model specification, including dynamical algorithm, see the SI.

\section{Discussion}

The main novelty of the approach in this article is to identify, in a large sample of countries, the existence of a systematic relationship between the appearance of new industries and the appearance or disappearance of other industries within the same country. 
In particular our analysis reveals that (i) products appear cluster-wise in creative bursts which consequently (ii) increases the chance for other, already existing products to be pushed from the market and (iii) the emerging products are typically associated with higher income and a higher level of economic complexity.
The effect of shifting production towards higher complexity is strongest in the developing countries of Latin America, Eastern Europe and East Asia and Pacific and barely visible in the least developed countries or Europe and North America (see SI).

These observations can be explained within an evolutionary model of Schumpeterian economic dynamics \cite{Thurner10}.
Just as the product portfolio of countries can be explained as a consequence of the presence of productive capabilities in that country \cite{Hausmann11a},
changes in a country's product portfolio can be understood through the evolution of its capabilities.
Entrepreneurs (firms) upgrade and recombine existing capabilities to create innovations.
These may substitute existing industries, restructuring the market.
Such a system can exist in three different modes \cite{htk1, htk2}.
First, if the initial number of capabilities is too low to allow for a sufficient number of novel combinations, the creative destruction process may not set in at all.
Second, if the country is already fully diversified progress is also slowed down.
In between these regimes there is a third, transition regime, where the creative destruction process is most effective.
We calibrate the model with initial product diversities and (dis)appearance rates observed through world trade data, and can reproduce (i) the position of the transition regime from low to high diversity, (ii) the patterns of (dis)appearance frequencies per country and (iii) the distribution of (dis)appearance frequencies for individual products.

Beyond finding evidence of creative destruction in a large set of national economies, 
this paper also helps to further reveal the complex topology of industry relatedness. 
Understanding

\section{Materials and methods}
\subsection{Surrogate data}
For the purpose of statistical analysis we construct a surrogate dataset with
the aim to destroy the correlations in the timing of (dis)appearances while the event statistics (as shown in Fig.\ref{Schump} (d) and (e)) are preserved.
The surrogate data is prepared as follows.
Each event is given by a triplet $(p_i,c_i,t_i)$ where index $i$ runs over all appearance (disappearance) events in the data. 
For each $i$ we fix $p_i$ and $c_i$ while shuffling the years $t_i$ between the events in the triplets.
Formally this defines a random permutation $\mathcal P: \{i\} \mapsto \{i\}$ over the set of all event indexes $i$.
One may then calculate the co-occurrence measures from the triplet $(p_i,c_i,\mathcal P(t_i))$ and average the result over many realizations of $\mathcal P$. This ensures that the marginal distributions of the events (Fig. \ref{Schump} (d) and (e)) remain unchanged while all correlations in the timing of appearances or disappearances are destroyed.

%Let $XY$ denote any of the four possible co-occurrences: appearance-appearance ($XY=AA$), appearance-disappearance ($AD$), disappearance-appearance ($DA$), or disappearance-disappearance ($DD$).
%$Q_{XY}(p,q,\tau)$ denotes the weighted co-occurrence measures for the surrogate data. 

\subsection{Network}
The creative bursts of cluster-wise co-appearing products can be effectively displayed as a network.
The pairwise conditional co-appearance measure, $\tilde P_{AA}(p,q)$, allows to construct similarity matrices for product categories from which the maximum spanning tree, shown in Fig. \ref{MST}. is constructed (see SI).
Nodes in the network represent product categories according to the SITC rev.2 4-digit classification. 
The color of the nodes represents the products' Leamer classes \cite{Leamer84}. This is a classification based on relative factor intensities such as amount of capital, labor, land, skills etc.
A similar route has been followed to construct the product space \cite{Hidalgo07}. 
Fig. \ref{MST} suggests that the clusters of co-appearing products require one or several common capabilities.
Once a country acquires or upgrades these capabilities, it starts to report exports in these product categories.
Their appearances are simultaneously observed as a creative burst within the same Leamer class.

\subsection{Product Indicators}

\emph{Product Income Indicator PRODY.} 
This indicator for a given product is the weighted average of the per capita GDPs of countries exporting it.
It is a weighted average of the income per capita of the countries that have revealed comparative advantage in that product \cite{Hausmann07}.
A more sophisticated product, in principle, should be made by richer countries.
Let us define the difference in $PRODY$ for products $p$ and $q$ as $\Delta PRODY(p,q,\tau) = $ $PRODY(p|$ appears at $t) - PRODY(q|$ disappears at $t')$ with $t<t'\leq t+\tau$.

\emph{Product Complexity Indicator.}
The  Product Complexity Index $PCI$ is an indicator for the economic complexity involved in manufacturing a given product \cite{Hausmann11}.
The $PCI$ is a combination of the ubiquity of a product (i.e. the number of countries that make it) and the economic diversity of its exporting countries, see SI for more details.
Products with high $PCI$ typically are made by few highly diversified countries which is indicative of high economic complexity \cite{Hidalgo09}.
The change in complexity $\Delta PCI(p,q,\tau)$ between two products is given by $\Delta PCI(p,q,\tau) = $ $PCI(p|$ appears at $t) - PCI(q|$ disappears at $t')$, $t<t'\leq t+\tau$.

\emph{Killer Index.}
We refer to $KI(p) \equiv SPI_{AD}^{(\tau)} (p)$ as the 'Killer Index' for each product $p$.
It measures the likelihood that the appearance of a given product $p$ is observed with the disappearance of any other product in the next $\tau$ years. 

\emph{Extinction Index.}
The likelihood that a product disappears together with the appearance of any other product $\tau$ years earlier is expressed by its 'Extinction Index'
$XI(q) =\sum_p P^{(\tau)}_{AD}(p,q) - P^{(\tau)}_{AD}(q,p)$. We set $\tau=3$.

\section{Acknowledgments}
We acknowledge extremely helpful discussions with Rudolf Hanel and financial support from FP7 project CRISIS.

\clearpage

\setcounter{figure}{0}

\huge 

\begin{center}
Supplementary Information for 'Empirical Confirmation of Creative Destruction'
\end{center}

\normalsize
\section{Description of Datasets}

We use two independent data sets in this work.
\begin{itemize}
\item {\bf NBER trade data} \cite{NBER}. Compiled by the National Bureau of Economic Research, this set of bilateral trade data by commodity spans the period 1962-2000. Trade flows (in USD) are reported in product categories following 4-digit SITC rev.2 classification. This dataset is a combination of two others, spanning 1962-1983 and 1984-2000 respectively. We work with the timespan 1984-2000 to exclude any possible artifacts in the results due to changes in data collection between these two timespans. The NBER trade data introduces artificial product categories (containing 'A's and 'X's in the SITC code) to account for differences in import and export records (i.e. if country A exports to countries B,C, but A's export record deviates from (B+C)'s import records). We only focus on export data and exclude these artificial product categories. Finally, we only include 'real' countries (the dataset also lists world regions, such as Southern Asia or Oceania, etc.). This results in longitudinal trade data for 200 countries in 800 product categories over 17 years.
\item {\bf COMTRADE trade data} \cite{COMTRADE}. The United Nations Commodity Trade Statistics Database (UN COMTRADE) publishes annual international trade statistics data by commodities and partner countries. We use data from the timespan 1990-2010. Export values (in USD) are reported in HS1992 product categories for over 170 countries (again, leaving aside world regions), amounting to roughly 5000 categories over 21 years.
\end{itemize}

Let $A(p,c,t)$ be a product indicator function for the appearance of product $p$ in country $c$ between year $t-1$ and $t$,

\begin{equation}
A(p,c,t) = \left\{
  \begin{array}{l l}
    1 & \quad \mathrm{if } \ x(p,c,t-1)=0 \ \mathrm{ and } \ x(p,c,t)>0\quad,\\
    0 & \quad \mathrm{otherwise}\quad.\\
  \end{array} 
 \right.
\label{AppEvSI}
\end{equation}

Similarly the indicator function for a disappearance event is
\begin{equation}
D(p,c,t) = \left\{
  \begin{array}{l l}
    1 & \quad \mathrm{if } \ x(p,c,t-1)>0 \ \mathrm{ and } \ x(p,c,t)=0\quad,\\
    0 & \quad \mathrm{otherwise}\quad.\\
  \end{array} 
 \right.
\label{DisAppEvSI}
\end{equation}

\begin{figure}[b]
\begin{center}
 \includegraphics[height=35mm]{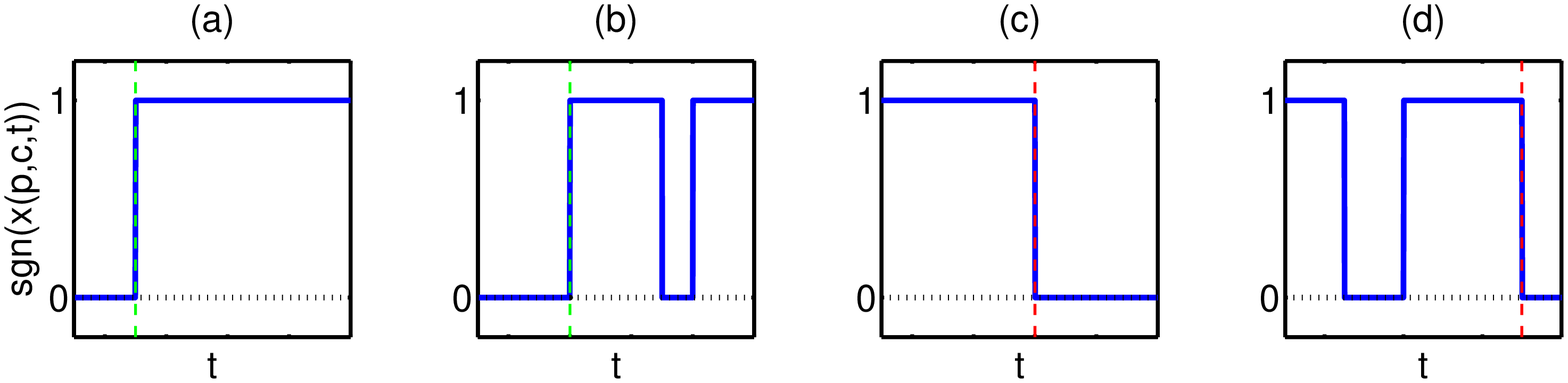}
  \end{center}
 \caption{Appearance and disappearance events in product $p$ in country $c$ are only included if the underlying diversity timeseries $\mathrm{sgn} (x(p,c,t))$ is of one of the types (a)-(d). The time of the appearance event is then always the first appearance (green dashed line in (a) and (b)), disappearances are the last events (red dashed lines in (c) and (d)).}
\label{DivSeries}
\end{figure}

Note that these definitions are only useful if there exists a data record for $c$ at both $t$ and $t-1$. 
We exclude small countries from the analysis by demanding a population of at least 1.2 million people and total exports of at least 1 billion USD, leaving us with a list of 125 countries.
The reported results for the SPI where computed over the timespan 1984-2000.
Individual trade flows between countries are only included if they exceed 100000 USD
Furthermore, appearance and disappearance events are {\bf not} included if one of the following is true.
\begin{itemize}
\item In the year before product $p$ appears (or after $p$ disappears) in a country $c$, the country reports positive exports in less than fifty different categories.
\item The diversity time-series $\mathrm{sgn} (x(p,c,t))$ of the product does not belong to one of the four types shown in Fig. S\ref{DivSeries}.
\end{itemize}
After these filtering procedures we assign each product a  appearance or disappearance event (or no event at all).
If existing, each product in each country is assigned a unique appearance (disappearance) event which is its first (last) measured event. For this particular event we set $A(p,c,t)=1$ ($D(p,c,t)=1$) and zero at each other point, in Fig. S\ref{DivSeries} appearance events are highlighted by green slashed lines, disappearances by red dashed lines.

\section{Robustness}

Several types of robustness checks are conducted. The analysis is repeated with a threshold of 200,000 USD for export values, that is whenever $x(p,c,t)<200000$ we set $x(p,c,t) \equiv 0$.
The results remain basically unaltered and are thus not driven by small, fluctuating export values, see column 'threshold' in Tab. 1.
The next check concerns artificial product appearances / disappearances  due to the creation or abandonment of product categories or countries. To this end only product categories are included in the analysis which have positive world exports in each year. This does not affect the analysis, see column 'all products with positive exports'.
Thirdly the influence of the time-span on the results was checked. The time-span was reduced to 1989-2000, see 'cleaned data 1989-2000'.
Former communist transition countries from the FSU are excluded in the column 'Excl. FSU'. Then the maximal time lag is varied to $\tau=5$ and $\tau=7$ years respectively.
Finally the same analysis is carried out with the UN ComTrade data.

\begin{figure}[tbp]
 \begin{center}
 \includegraphics[height=120mm]{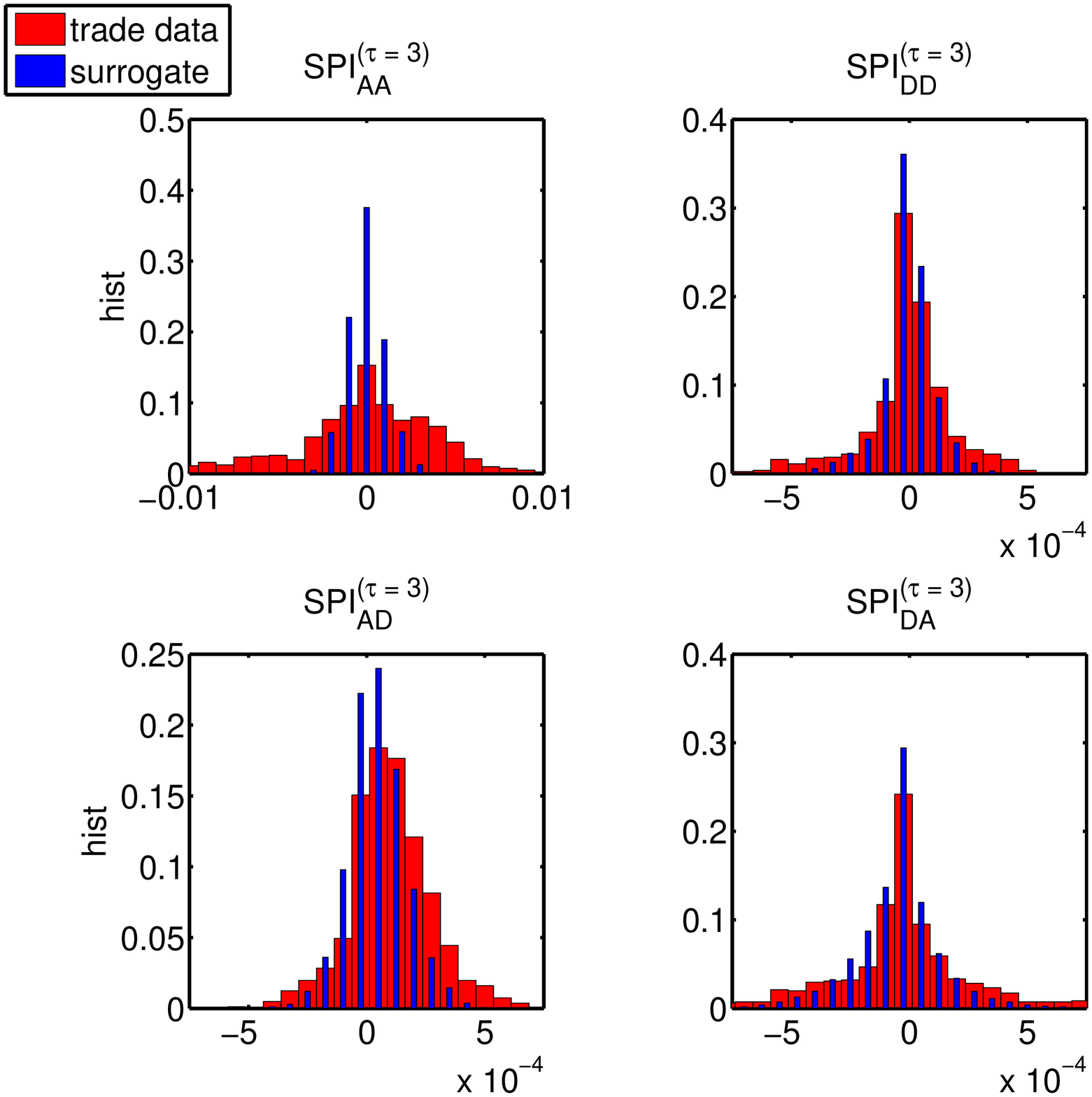}
  \end{center}
 \caption{Histograms for Schumpeterian Product Index $SPI^{(\tau)}_{XY}(p)$ for co-appearances $AA$, co-disappearances $DD$, and the mixed forms $AD$ and $DA$ (from top to bottom) for maximal time-lag $\tau=3$. The $SPI$ is shown for the surrogate data (blue) and the trade data (red). A significant fraction of higher SPI values for the trade data are seen for appearance-disappearances.}
 \label{SPI_Hists}
\end{figure}

\section{Maximum Spanning Tree}
The MST is the network with the least number of links with the highest possible weights spanning each node of the network. 
It can be computed using Kruskal's algorithm. The MST can be regarded as the 'skeleton' or 'backbone' of a network.
It positions each node in the neighborhood of other nodes to which it has 
strong ties and is therefore reminiscent of hierarchical clustering procedures.

\section{Result for COMTRADE database}

All results reported in the main text can also be found in the COMTRADE database. We have chosen to work with the NBER data for the following reasons.
The product classification employed for COMTRADE has gone through several revisions, namely HS1992, HS1996 and HS2002. 
Each of these revisions causes artificial appearances and disappearances (due to re-classification of products).
At the time the dataset was extracted a substantial amount of trade data was still to be reported causing artificial disappearances, thus one can only effectively work with the range 2000-2009.
The NBER database, on the other hand, is the outcome a collaborative research effort to provide a coherent dataset. 
For example, if a country only reports its trade flows in 3-digit SITC or an older revision, this is painstakingly checked against the reports of the trade partners and records from the US trade database to 'fill the gaps'.
This reduces the number of artificial appearances and disappearances. A detailed description of how the NBER data was composed can be found in \cite{NBER}.
$p$-Values for the comparison of trade and surrogate data are shown in Tab. 1, column 'COMTRADE'

\section{Further results on product comparisons}

\begin{figure}[tbp]
  \begin{center}
  \includegraphics[height=60mm]{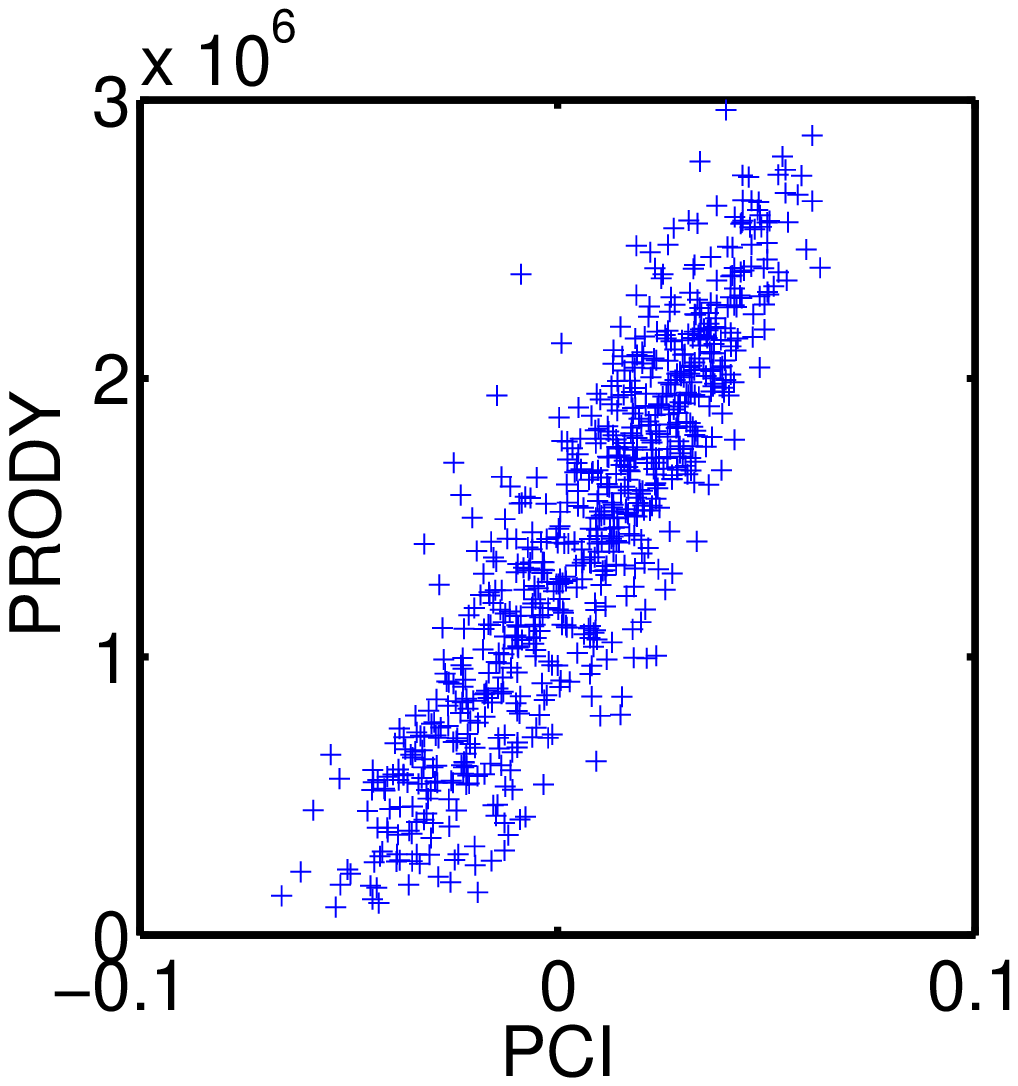}
   \end{center}
  \caption{Comparison of PRODY and PCI values for each product category.}
  \label{PRODY_PCI}
 \end{figure}

The PRODY value was computed as in \cite{Hausmann07} using NBER trade data as an average over each PRODY from the timespan 1993-2000.
To compute the PCI we extract the matrix of significant exporters $M_{cp}$ as defined in \cite{Hidalgo09} from the NBER trade data for each year from 1993-2000.
Using $k_{c,0} = \sum_p M_{cp}$ and $k_{p,0} = \sum_c M_{cp}$ one computes the matrix $\tilde M_{pq} = \sum_c \frac{M_{cp} M_{cq}}{k_{c,0} k_{p,0}}$. The PCI is the average over all years of the eigenvector associated with the second largest eigenvalue of $\tilde M_{pq}$. Fig.\ref{PRODY_PCI} shows how these two indexes are related.

\section{Further results on country comparisons}

\begin{figure}[tbp]
 \begin{center}
 \includegraphics[height=55mm]{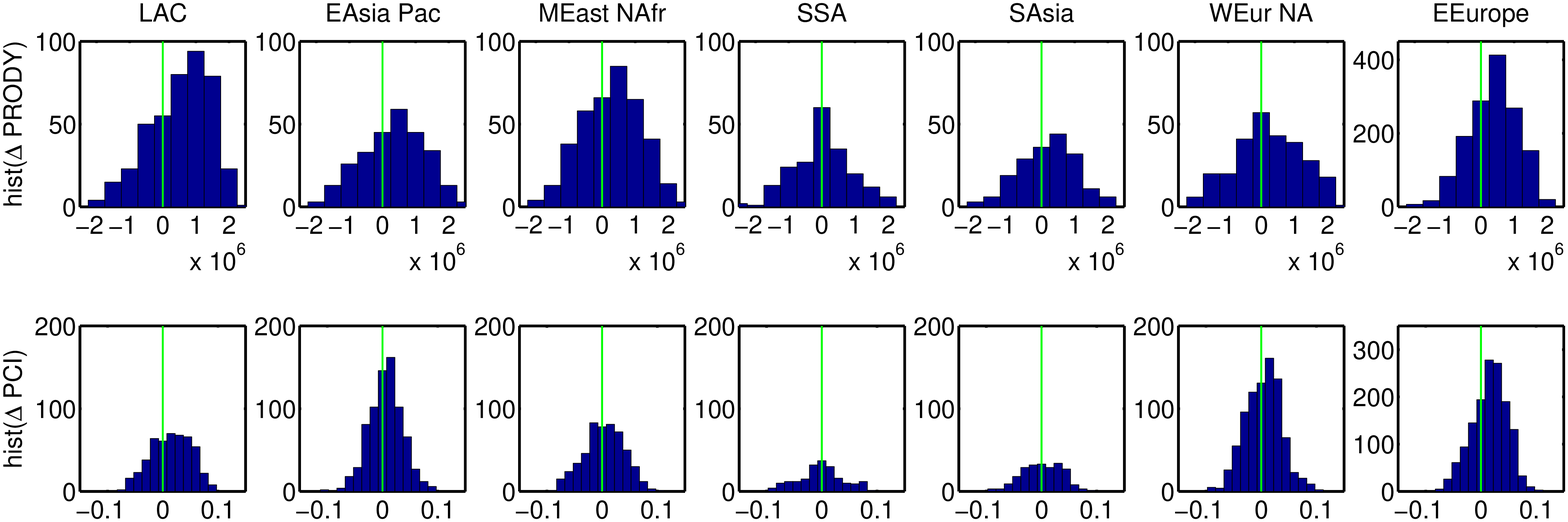}
  \end{center}
 \caption{Histograms for change in (top row) income level $PRODY$ and (bottom row) product complexity $PCI$ associated with each $AD$ process for seven different world regions (Latin America and the Carribean, East-Asia and the Pacific, Middle-east and North-Africa, Sub-saharan Africa, South-Asia, Western Europe and Northern America, Eastern Europe). There is a skew to the positive side most clearly visible for Latin America and the Carribean, as well as for CEE countries.}
 \label{DPCI_Regions}
\end{figure}

Not each region in the world shows the same pattern of development towards higher economic complexity. We repeat the analysis of complexity change associated with $AD$ processes for countries from Latin America and the Carribean, East-Asia and the Pacific, Middle-east and North-Africa, Sub-saharan Africa, South-Asia, Western Europe and Northern America, as well as Eastern Europe separately.
That is, we use only countries from the respective regions, compute $SPI_{AD}^{(\tau)}(p)$ for these countries and compute $\Delta PRODY$ and $\Delta PCI$ as described in the main text, see Fig. S\ref{DPCI_Regions}.
Transitions towards higher economic complexity (skew to the right in the histograms) is clearly visible for Latin America and Eastern Europe. In other regions this trend is less visible or even absent.
This supports the notion of 'critical economies' where Schumpeterian economic evolution is successful at work. These critical economies can be identified as transition countries. 
For low diversity countries (compare e.g. Sub-saharan Africa) there is simply 'too little to start with' in order to move towards higher economic complexity, with respect to this they can be called 'sub-critical'.
Fully diversified countries (compare Western Europe and North America) also defy this trend. Since they are already at the top of the 'complexity scale', there is not much room for improvement here.
This may be due to an too coarse resolution of the standard product classification. These countries can be dubbed supra-critical.

\begin{figure}[tbp]
 \begin{center}
 \includegraphics[height=100mm]{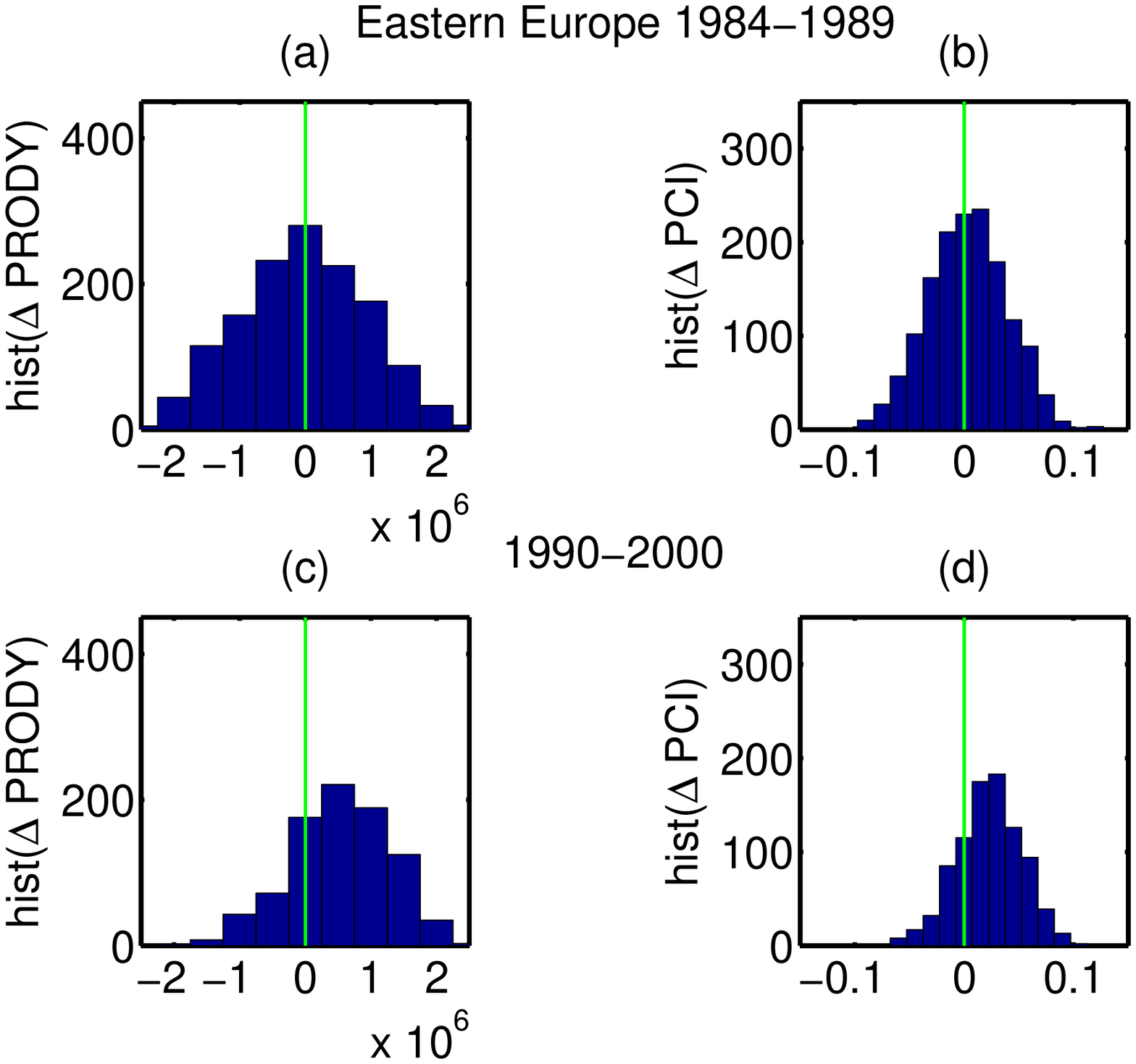}
  \end{center}
 \caption{Histograms for change in income level $PRODY$ and (a) product complexity $PCI$ (b) associated with each $AD$ process for Eastern European countries between 1984-1989, that is in a centrally planned economy. The development is more or less symmetric. Compare this to the clear positive skew in (c) and (d), where changes in $PRODY$ and $PCI$ for the same countries between 1990-2000 are shown. This suggests that the increase in economy complexity observed there is due to the shift from planned to market economy.}
 \label{Hist_EE}
\end{figure}

Eastern European countries have undergone a major shift in their political regimes in the timespan we use for the measurement. One may wonder whether the shift towards higher economic complexity is due to developments under the centrally planned economy before 1990 or afterwards. The analysis is repeated for these countries using the timespan 1984-1989, see Figs. S\ref{Hist_EE}(a) and (b).
Compare these histograms to the ones observed between 1990-2000 in Figs. S\ref{Hist_EE}(c) and (d). There is no clear movement towards higher complexity before 1990. This suggests that the move towards higher complexity observed later is due to the shift from a centrally planned to a free market economy.

In analogy to the SPI a Schumpeterian Country Index SCI is defined  for $AD$ and $\tau = 3$ as
\begin{equation}
SCI(c) = \sum_{p,q,t,t'} \left( A(p,c,t) D(q,c,t') - A(q,c,t) D(p,c,t') \right) \quad t<t'\leq t+\tau
\quad.
\label{SCI}
\end{equation}

\begin{figure}[tbp]
  \begin{center}
  \includegraphics[height=80mm]{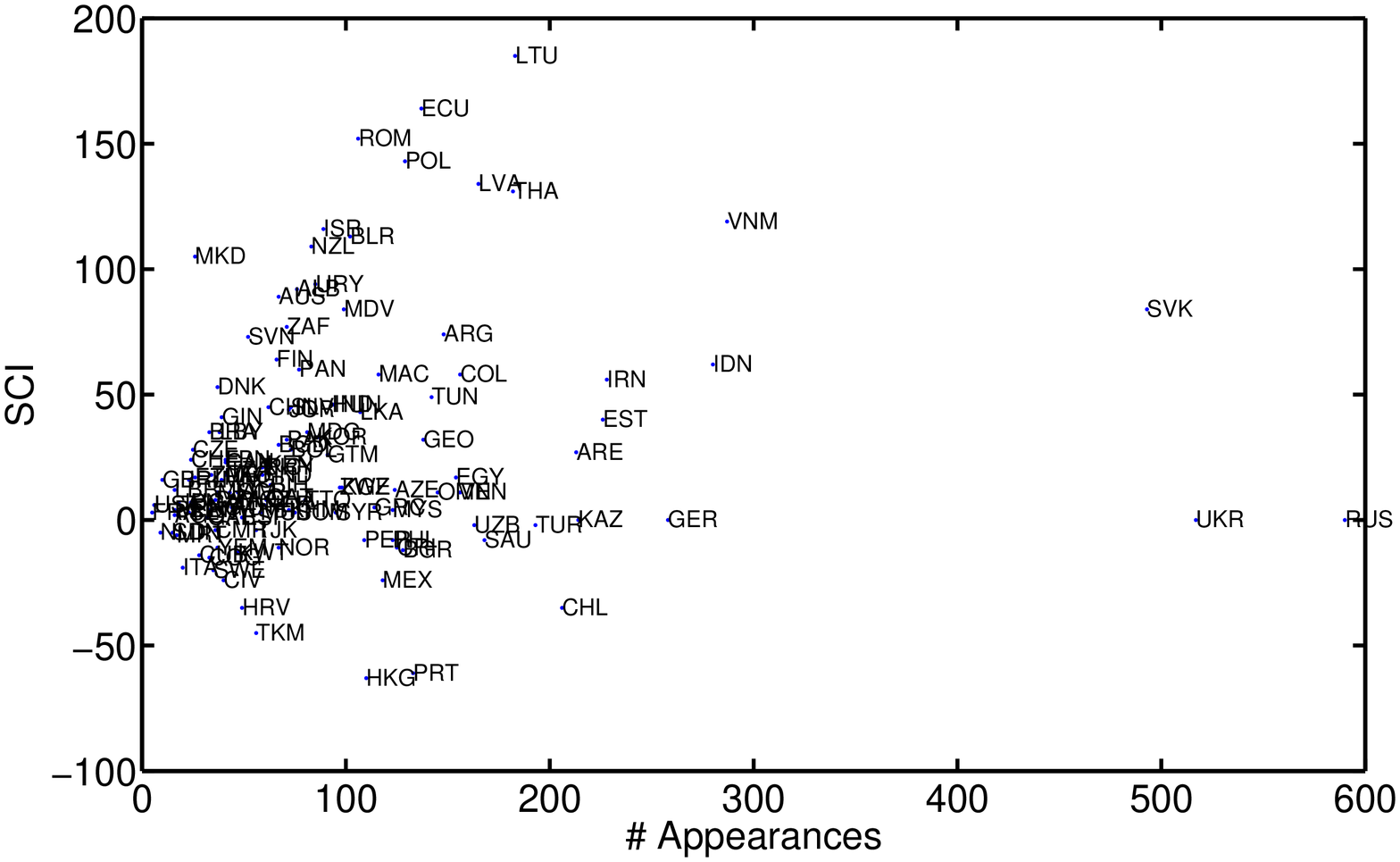}
   \end{center}
  \caption{Number of appearances vs SCI for each country. This shows that SCI can not be explained by the number of appearances alone. Observe that e.g. Ecuador and Portugal have the same number of product appearances, but Ecuador has one of the highest SCIs and Portugal one of the lowest.}
  \label{AppsSCI}
 \end{figure}

We show that the SCI measure is not an artifact of the number of appearances alone in Fig. S\ref{AppsSCI}, where the SCI for each country is plotted against the number of appearances in this country.
Countries like Portugal and Hongkong have one of the lowest ranking SCIs, Ecuador and Lithuania one of the highest ranking ones. But each of these countries have rougly the same number of product appearances.

\begin{figure}[tbp]
  \begin{center}
  \includegraphics[height=80mm]{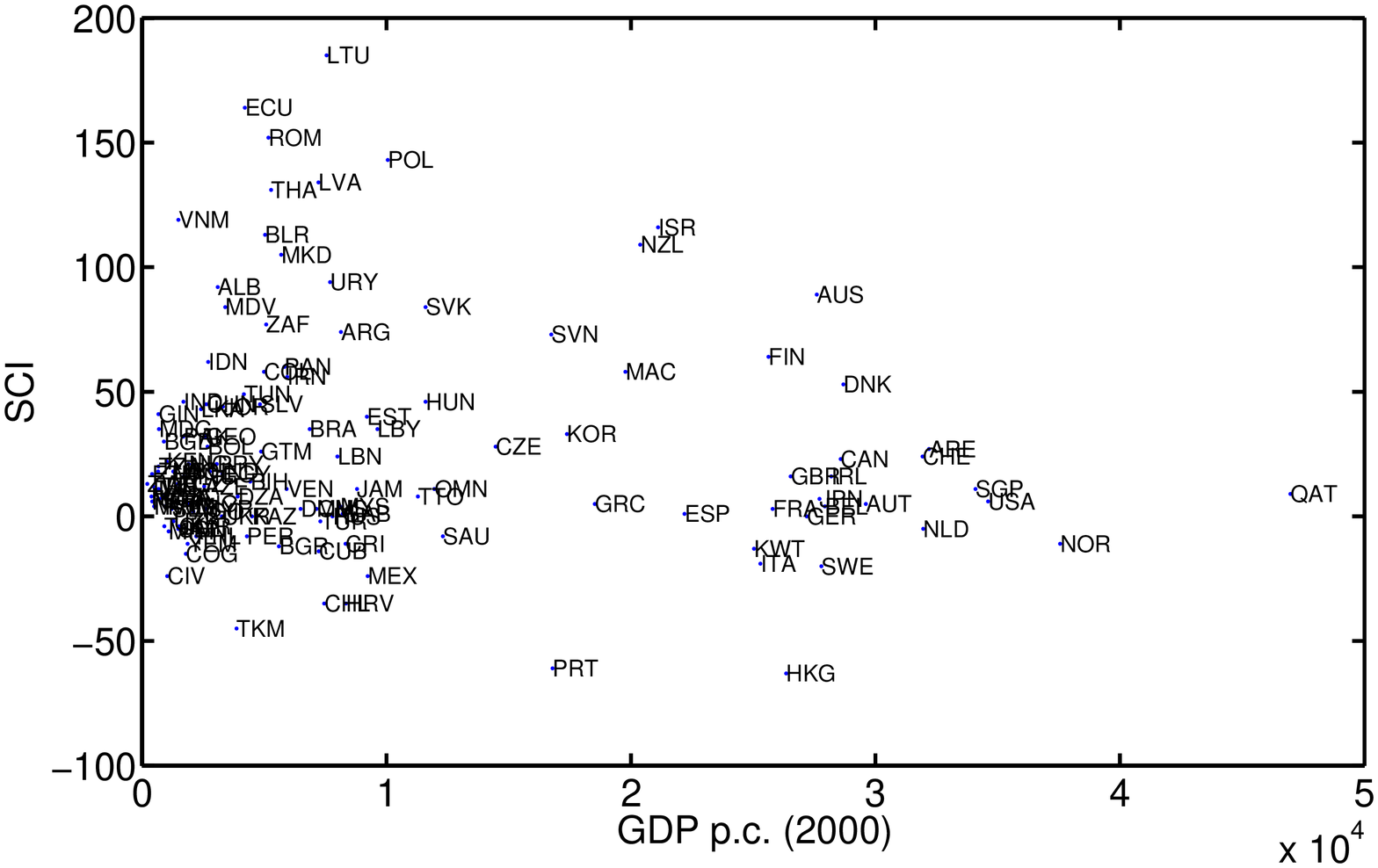}
   \end{center}
  \caption{GDP vs SCI for each country. Countries cluster into two distinct groups. Western countries are mostly found at the bottom right (low SCI and high GDP), poor countries to the bottom left (low SCI and low GDP), countries with high SCI form a bridge between those two regimes.}
  \label{GDP_SCI}
 \end{figure}

Countries with lowest ranking SCI values are countries which are either already fully diversified and show thus only little activity in terms of (dis)appearances (USA, Germany, France, China, UK, ...)
or countries with constant low diversities (Malawi, Gabon, Uganda, ...).
Countries with a relative high SCI include Uruguay, Kuwait, Croatia, ... and tend to be bridge countries between these two regimes.
Fig. S\ref{GDP_SCI} shows countries and their SCI and Gross Domestic Product (GDP) per capita of 2000. Western countries are mostly found at the bottom right (low SCI and high GDP), poor countries to the bottom left (low SCI and low GDP), countries with high SCI form a bridge between those two regimes.

\begin{figure}[tbp]
  \begin{center}
  \includegraphics[height=80mm]{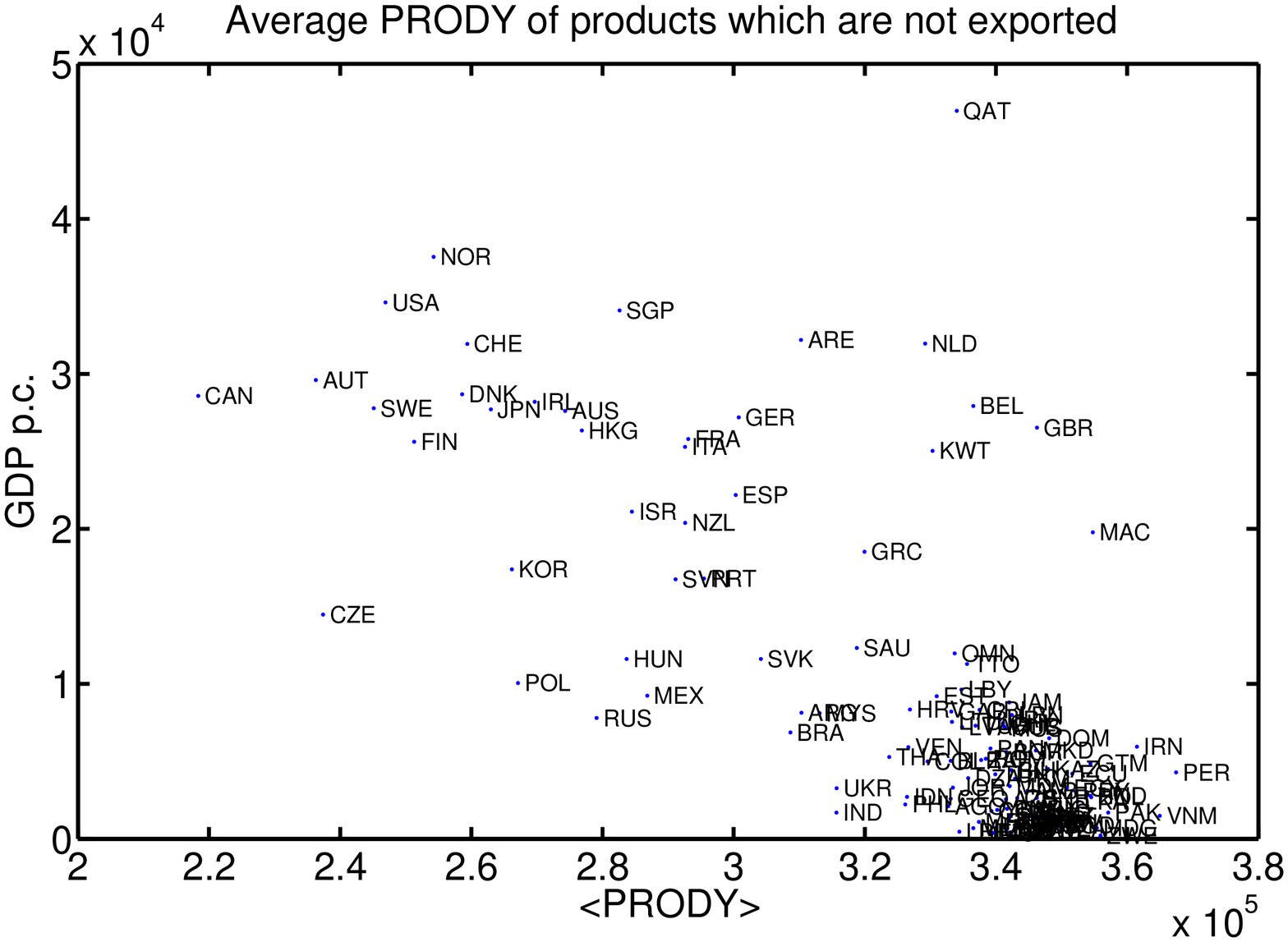}
   \end{center}
  \caption{GDP vs average PRODY of products not being exported by this country. High GDP countries tend to {\it not} export products with a low PRODY, i.e. low economic complexity (upper left region of the plot). Low GDP countries miss products with a high PRODY in their export baskets, they are found in the lower right region. Oil exporting countries such as Qatar and the United Arabian Emirates deviate from this trend. They achieve high GDP per capita values with significantly less complex products.}
  \label{CompPRODY}
 \end{figure}

The virtues of having or not having a product are shown in Fig. S\ref{CompPRODY}. For each country we compute the average PRODY value of products which are {\it not} exported by this country as of 2000.
High GDP countries (compare Canada, Austria, USA, Norway, Sweden, ...) do {\it not} export products which have on average a significant lower PRODY than developing and least developed countries.
There are some outliers, for example oil exporting countries achieving a high GDP per capita with low complexity products. 

\begin{figure}[tbp]
 \begin{center}
 \includegraphics[height=120mm]{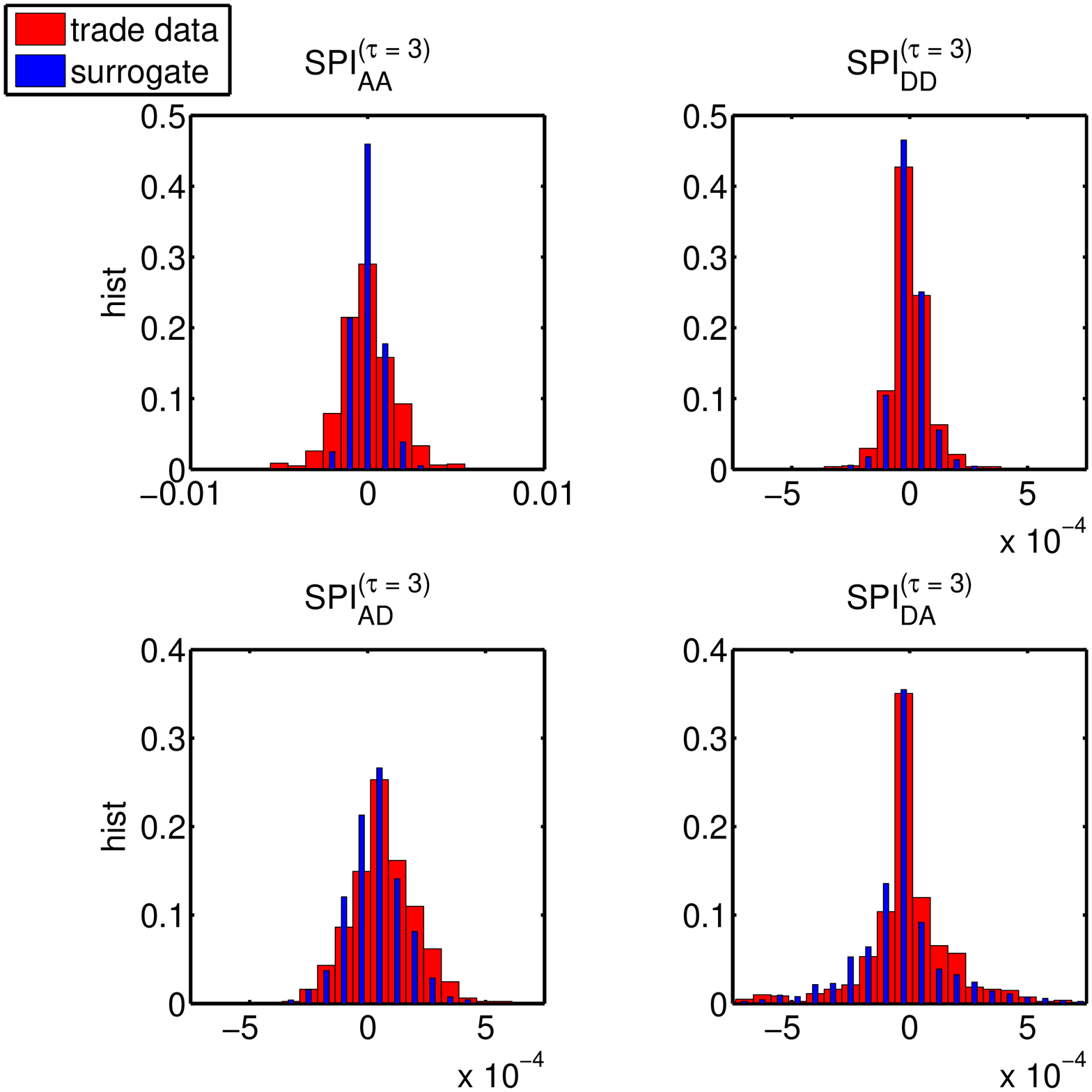}
  \end{center}
 \caption{Histograms for Schumpeterian Product Index $SPI^{(\tau)}_{XY}(p)$ for co-appearances $AA$, co-disappearances $DD$, and the mixed forms $AD$ and $DA$ (from top to bottom) for maximal time-lag $\tau=3$. Countries from the Former Soviet Union and Eastern Europe are excluded. The $SPI$ is shown for the surrogate data (blue) and the trade data (red). A significant fraction of higher SPI values for the trade data are seen for appearance-disappearances.}
 \label{SPI_Hists_ExclFSU}
\end{figure}

Since the transition countries from the Former Soviet Union are main contributors to the $AD$ process, as can be observed from the previous results shown in Fig. S\ref{DPCI_Regions}, Fig. S\ref{AppsSCI} and Fig. S\ref{GDP_SCI}, it is worth to show that the significance of the $AD$ results is not due to the contributions of these countries alone. 
To this end we compute histograms for $SPI^{\tau}_{XY}(p)$ for a set of countries excluding the FSU and Eastern Europe, 
see Fig. S\ref{SPI_Hists_ExclFSU} and the respective column of $p$-Values in Tab. 1 entitled 'Excl. FSU'.

\section{Schumpeterian Evolutionary Model}

The occurrence of bursts of creation (product appearance) and destruction (product disappearance) is a prominent feature of Schumpeterian economic dynamics \cite{Thurner10}. We show that this model can be used to describe the development of national economies in terms of product (dis)appearances as Schumpeterian 'gales of destruction'. The model is a variant of \cite{Thurner10} with some features proposed by \cite{Hidalgo09}. The components of the model are now described in detail.

\subsubsection{Model Components}
{\bf Products.}
For each country $c$ the binary product vector $\pi_{c,p}(t)$ records whether product $p$ is currently produced ($\pi_{c,p}(t)=1$) or not ($\pi_{c,p}(t)=0$). To produce $p$ it is required to possess a specific set of capabilities.

{\bf Capabilities.}
A country's product portfolio is determined by the available nontradable capabilities residing in it \cite{Hidalgo09}. Each country $c$'s set of capabilities is described be a time-dependent vector $\sigma_c(t)$ with binary components. The $i$'th entry in $\sigma_c(t)$ indicates whether the capability $i$ is available in country $c$ at time $t$ ($\sigma_{c,i}(t) = 1$) or is not available ($\sigma_{c,i}(t)=0$). The total number of capabilities which can be acquired is fixed to some $N_A \gg 1$.

{\bf Productions - Capability Gain.}
Capabilities are acquired through a novel combination of already present capabilities. This is encoded in the production rule table $\alpha^+_{ijk}$. If capability $i$ can in principle be acquired through a combination of $j$ and $k$ then $\alpha^+_{ijk} = 1$, otherwise $\alpha^+_{ijk}=0$. Note that each the production rule table is the same for each country. The production process is then given by
\begin{equation}
\sigma_{c,i}(t+1) = \alpha^+_{ijk} \sigma_{c,j}(t) \sigma_{c,k}(t)
\quad.
\label{ProdProc}
\end{equation}
There are on average $r^+$ unique and different production processes for each capability $i$, with $j$ and $k$ chosen at random from the entire set of capabilities.

{\bf Product disappearance.}
If a product requires a set of capabilities which is a subset of capabilities required to produce another product, it may disappear.
The idea is that the product with the larger set of capabilities is a more complex and advanced product substituting its antecedent (e.g. wikipedia replacing an encyclopedia). 
Assume product $p$ requiring capability $i$ and product $q$ requires capability $j$. If there is a production rule $\alpha^+_{ijk}$ then product $p$ requires capability $j$ {\it and} $k$ and is thus an improved version of product $q$. We incorporate this as competitive replacement in the model by assuming that with probability $p^-$ a production rule for $i$ leads to a negative influence on $j$.
For this the destruction rule table $\alpha^-_{ij}$ is introduced. It encodes whether capability $i$ leads to a substitution for $j$ ($\alpha^-_{ij}=1$) or not ($\alpha^-_{ij}=0$). The destruction process is given by
\begin{equation}
\sigma_{c,i}(t+1) = 1- \alpha^-_{ij} \sigma_{c,j}(t) 
\quad.
\label{DestProc}
\end{equation}
The destruction rule table is ubiquitous for each country.

{\bf Migration of Capabilities.}
In addition to the deterministic production and substitution processes there is a stochastic migration of capabilities among the countries. With probability $p$ a capability migrates from a country $c_1$ to another, randomly chosen country $c_2$. If $\sigma_{c_1,i}(t) = 1$ we set $\sigma_{c_1,i}(t) = 0$ and $\sigma_{c_2,i}(t) = 1$ with probability $p$.

{\bf From Capabilities to Products.}
Each product needs $n_a$ different capabilities as input. This is accounted for in the capability $\times$ product matrix $M_{ip}$. If product $p$ requires capability $i$ then $M_{ip}=1$, otherwise $M_{ip}=0$. Thus each column has $n_a$ randomly distributed nonzero entries. In each timestep it is checked for each product whether its set of required capabilities is present. Thus $\pi_{c,p}(t)=1$ if and only if $\sum_i M_{ip} \sigma_{c,i}(t) = n_a$.

\subsubsection{Dynamical Algorithm}

The Schumpeterian dynamical algorithm for this model is given by:
\begin{itemize}
\item Pick a country $c$ at random.
\item For each capability $i$ with $\sigma_{c,i}(t)=1$, set $\sigma_{c,i}(t)=0$ and $\sigma_{c',i}(t)=1$ with probability $p$ (where $c'$ is another randomly chosen country).
\item Pick a capability $i$ at random.
\item Sum all productive and destructive influences on $i$, that is compute $\Delta_i(t) \equiv \sum_{j,k} \alpha^+_{ijk} \sigma_{c,j}(t) \sigma_{c,k}(t) - \sum_j \alpha^-_{ij} \sigma_{c,j}(t)$. If $\Delta_i(t)>0$ ($\Delta_i(t)<0$) then $\sigma_{c,i}(t+1) = 1$ ($\sigma_{c,i}(t+1) = 0$), otherwise $\sigma_{c,i}(t+1) = \sigma_{c,i}(t)$.
\item Continue picking capabilities until all capabilities have been updated, then go to the next country.
\item Once all countries have been updated, update the product vectors $\pi_{c,p}(t)$ and go to the next timestep.
\end{itemize}

The model is initialized with a number of countries $N_c$ and products $N_p$ taken from the data. As further model input serves the number of products that each country exports at the initial time-step $t=0$ (corresponding to 1984 for the data). Let this diversity be $D_c(0)$. We set $\sigma_{c,i}(0) = 1$ with probability $(D_c(0)/N_p)^{1/n_a}$. This guarantees that the initial model diversity $\sum_p \pi_{c,p}(0)$ equals $D_c(0)$. After initialization, the model is iterated over $T'$ time-steps. Assume that we compare the model to an empirical timeseries over $T$ years. The resulting model trajectories are segmented into $T$ segments of equal length ('model years'). The value of $T'$ is chosen such that the absolute number of appearance and disappearance events is the same in model and real world data. This rescaled model data is now called $\bar x(p,c,t)$ and can be compared to the real world data $x(p,c,t)$. We will denote model entities by a bar, e.g. $\bar A(p,c,t)$ denotes appearance events obtained from $\bar x(p,c,t)$ via the same procedure as for $x(p,c,t)$.

The results of a comparison between the Schumpeterian diversity dynamics model and world trade data are summarized in Fig. \ref{Schump}. 
Here the model parameter $r^+=1.65$, $p^- = 0.15$, $p=0.002$, $N_A = 100, n_a=2$. All other model parameters can be measured in the data. 
In Figs.\ref{Schump}(a),(b) the initial and final product diversities for each country are shown. In both cases the countries are ranked by diversity. Per calibration the initial diversities are equal. The final diversities (b) also overlap to a large extent.
A more detailed view of this is given in Fig. \ref{Schump}(c). Here the increase in product exports for countries of a given rank is shown. A remarkable peak is visible in both model and trade data.
In the model this is explained by the onset of a creative phase transition \cite{htk1},\cite{htk2}. 
As a country gradually acquires novel capabilities it eventually reaches a point where it has accumulated enough critical capabilities to diffuse into the entire product space.
Creative bursts of diversification are the consequence.
Countries below this critical point experience stagnation, whereas countries above this threshold reach a fully diversified plateau.

The results reported are independent of $N_A$ \cite{htk1}.
The variable $p$ merely sets the time-scale for $T'$.
The parameters $r^+$ and $p^-$ determine the number of capabilities required for creative bursts to set in, that is the phase transition to a fully diversified country.
Changing these values rescales the peak in Fig.\ref{Schump}(d). The main conclusions however remain intact.

\end{document}